\documentclass[onecolumn,12pt,aps,prd,nobibnotes,showpacs,showkeys,
unsortedaddress]{revtex4-2}
\usepackage{graphicx,graphics,color,epsfig}
\usepackage[colorlinks=true,linkcolor=blue,citecolor=blue,urlcolor=blue,
breaklinks=true]{hyperref}
\usepackage{amsmath,amsfonts,amssymb}
\usepackage{verbatim}
\usepackage{latexsym}
\usepackage{dcolumn}
\usepackage{bm}
\usepackage{natbib}

\begin{document}

\title{\href{}{Active and sterile neutrino oscillations inside the Sun\\
in a phenomenological $(3+1+2)$-model}}
\author{V.~V. Khruschov}
\email{khruschov{\_}vv@nrcki.ru}
\affiliation{National Research Center ``Kurchatov Institute'',
Kurchatov~place~1, 123182~Moscow, Russia}
\author{S.~V. Fomichev}
\email{fomichev{\_}sv@nrcki.ru}
\affiliation{National Research Center ``Kurchatov Institute'',
Kurchatov~place~1, 123182~Moscow, Russia}

\begin{abstract}
The phenomenological model with three active and three light sterile neutrinos is considered taking into account terrestrial experimental
data, which indicate anomalies at short distances beyond the minimally modified Standard Model with three massive active neutrinos. One of the
sterile neutrinos is assumed to have distinctly different mass in comparison with masses of two others, that is corresponding to a $(3+1+2)$-model of neutrinos. Model parameters values used for the
description of oscillations of both active and sterile massive neutrinos
into the Sun are chosen. Oscillation characteristics of solar neutrinos
together with sterile neutrinos contributions have been evaluated taking
into account the neutrino interaction with the matter inside the Sun
with the use of the standard solar model (SSM). Results obtained
correlate with observational data and can be used for development of
sterile neutrinos models.
\end{abstract}



\maketitle

\section{Introduction}

Investigation of spectral characteristics of neutrinos emitted from the Sun
and Supernovae is among methods used for adequate description of processes
occurring during the star evolution, as well as for understanding the
fundamental properties of neutrinos. For instance, the solution of the solar
neutrino deficit problem with the Mikheyev--Smirnov--Wolfenstein (MSW) effect resulted in experimental discovery of neutrino oscillations and confirmation of the Standard Solar Model (SSM) as well as nonzero neutrino masses. Determination of values of neutrino masses and mixing parameters, as well as
a number and nature of different types of neutrinos are still the open
problems of the neutrino physics. As is expected, the final solution of these problems will be given taking into account experimental data concerning with a Grand Unification Theory (GUT), for which a generally accepted version is not currently available \cite{1}. To make a search for a requisite direction of extension of the Standard Model (SM) of the fundamental interactions to a certain GUT the various phenomenological models of neutrino are also under investigations. In this process the obtained and expected results of both carried out and planned experiments such as KATRIN, BOREXINO, DoubleCHOOZ, SuperNEMO, KamLAND--Zen, EXO, IceCube, DANSS, BEST, etc. are of a great importance.

Among the obtained experimental results about neutrino oscillations \cite{2}, one can refer indications on presence of anomalies for the neutrino and
antineutrino fluxes in different processes at short distances, which cannot
be explained by oscillations of three active neutrinos \cite{3}. These
anomalies may be associated with existence of sterile neutrinos and under
its confirmation they will be beyond not only the SM but also the minimally modified Standard Model ($\nu$SM) with three active massive neutrinos. The terrestrial short-baseline (SBL) neutrino experiments indicate a possible value of the sterile neutrino mass scale about 1\,eV \cite{4}. Besides, observational data related to formation of galaxies and their clusters
can be explained if one supposes existence of sterile neutrinos with masses
of the order of a few keV \cite{5,5a}. In the paper \cite{6}, it is shown
that the inclusion of the sterile neutrino with a mass about 0.5\,eV can resolve the tension between the Planck and BICEP2 results. More details on
possible existence of sterile neutrinos and their characteristics can be
found in numerous papers (see, e.g., Refs.~\cite{7,8,9,10,10a}). In the present paper we restrict ourselves to sterile neutrinos with masses of the
order of 1\,eV or lighter, which represent the interest for explanation
of the indicated SBL anomalies for the electron neutrino and antineutrino
fluxes as well as for oscillations of active neutrinos inside the Sun \cite{6,7,10a,11,12}.

It is well known that in general case there is no a limit to a number of sterile neutrinos, which do not interact with the known fundamental particles. At present, the most popular phenomenological schemes with both active and sterile neutrinos are $(3+1)$- and $(3+2)$-models \cite{13} with one or two sterile neutrinos, so the $\nu$SM can be called as the $(3+0)$-model. In principle, it is sufficient to introduce only one or two sterile neutrinos for description of the experimentally observed now anomalies for the neutrino fluxes. On the other hand, if one takes into account possible existence of the left--right symmetry of the weak and superweak interactions and associates sterile neutrinos with right neutrinos, one can conjecture that three sterile neutrinos are allowable \cite{14,14a,14b,15,16,17,18,19}, so it is desirable to consider the $(3+3)$-models. For example, the model with three light sterile neutrinos for the description of solar neutrino oscillations was considered in Ref.~\cite{20}. In the present paper we also consider the effect of three sterile neutrinos on survival and appearance probabilities of neutrinos emitted from the Sun. This is performed in the framework of the phenomenological $(3+1+2)$-model with three sterile
neutrinos with masses of the order of 1\,eV or less, when a mass value of
one of sterile neutrinos distinctly differs from mass values of two others (see, e.g., \cite{14a,14,14b}). This model is attractive both from the theoretical point of view and due to the good compatibility with available data from terrestrial experiments \cite{15}. Nevertheless, there exist the known problems of matching the number of light sterile neutrinos with the cosmic microwave background data \cite{21,22,25,26,26a}, and the resolution of this issue depends on applied cosmological models. For instance,
it was shown in Ref.~\cite{27}, if the peak of production rate of sterile
neutrinos occurs during a non-standard cosmological phase, the number density of relic sterile neutrinos could be significantly reduced.

\section{Oscillations Characteristics of Active Neutrinos}

As is well known, the oscillations of the solar, atmospheric, reactor and
accelerator neutrinos can be explained by mixing of neutrino states with
different masses. It means that the flavor states of neutrino are mixtures,
at least, of three neutrino mass states, and vice versa. The active neutrinos mixing is generally described by the $3\times3$ Pontecorvo--Maki--Nakagawa--Sakata matrix $U_{PMNS}\equiv U=V\times P$ with
the relations $\psi_{\alpha L}=U_{\alpha i}\psi_{iL}$, where
$\psi_{\alpha L}$ and $\psi_{iL}$ are the left chiral flavor and mass neutrino states, respectively, with $\alpha=\{e,\mu,\tau\}$ and $i=\{1,2,3\}$, and summation over repeated indices is implied. The matrix
$V$ can be written in the standard parametrization \cite{1} as
\begin{equation}
V=\left(\begin{array}{*{20}c}
1\hfill & 0\hfill & 0\hfill\\
0\hfill & c_{23}\hfill & s_{23}\hfill\\
0\hfill & -s_{23}\hfill & c_{23}\hfill\\
\end{array}\right)\times\left(\begin{array}{*{20}c}
c_{13}\hfill & 0\hfill & s_{13}e^{-i\delta_{CP}}\hfill\\
0\hfill & 1\hfill & 0\hfill\\
-s_{13}e^{i\delta_{CP}}\hfill & 0\hfill & c_{13}\hfill\\
\end{array}\right)
\times\left(\begin{array}{*{20}c}
c_{12}\hfill & s_{12}\hfill & 0\hfill\\
-s_{12}\hfill & c_{12}\hfill & 0\hfill\\
0\hfill & 0\hfill & 1\hfill\\
\end{array}\right),
\label{eq1}
\end{equation}
through the quantities $c_{ij}\equiv\cos\theta_{ij}$ and
$s_{ij}\equiv\sin\theta_{ij}$, and with the phase $\delta_{CP}$ (the Dirac
phase) associated with the CP violation in the lepton sector. The $3\times3$
matrix $P$ is the diagonal one,
$P={\rm diag}\{e^{i\alpha_{CP}},e^{i\beta_{CP}},1\}$, with $\alpha_{CP}$ and
$\beta_{CP}$ the additional CP-violating phases (the Majorana phases), which
cannot be identified with neutrino oscillations experiments.

The experimental data obtained in the neutrino oscillations experiments give
evidence of violation of the conservation laws for the leptonic numbers
$L_{e}$, $L_{\mu}$ and $L_{\tau}$. Besides they clearly point out to existence at least of two nonzero and different neutrino masses, by virtue of deviation from zero of two oscillation parameters $\Delta m_{12}^2$ and $\Delta m_{13}^2$ (with $\Delta m_{ij}^2=m_i^2-m_j^2$, where $m_{i}$ are the neutrino masses). Below we present the experimental values of oscillation parameters, which determine three-flavor oscillations of active neutrinos.
Together with the standard uncertainties on the level of $1\sigma$, these data obtained as a result of a global analysis of sufficiently high-precision measurements are as follows \cite{2}:
\begin{subequations}
\begin{align}
&\hspace{5cm}\sin^2\theta_{12}=0.304_{-0.012}^{+0.013}\,,\\
&\sin^2\theta_{23}=\left\{{{\begin{array}{*{20}c}
{\mbox{NH}:\quad 0.573_{-0.020}^{+0.016}}\hfill \\
{\mbox{\,\,IH}:\quad 0.575_{-0.019}^{+0.016}}\hfill \\
\end{array}}}\right., \quad
\sin^2\theta_{13}=\left\{{{\begin{array}{*{20}c}
{\mbox{NH}:\quad 0.02219_{-0.00063}^{+0.00062}}\hfill \\
{\mbox{\,\,IH}:\quad 0.02238_{-0.00062}^{+0.00063}}\hfill \\
\end{array}}}\right., \\
&\hspace{5cm}\delta_{CP}/\circ=\left\{{{\begin{array}{*{20}c}
{\mbox{NH}:\quad 197_{-24}^{+27}}\hfill \\
{\mbox{\,\,IH}:\quad 282_{-30}^{+26}}\hfill \\
\end{array}}}\right., \\
&\Delta m_{21}^2/10^{-5}\mbox{eV}^2=7.42_{-0.20}^{+0.21}\,, \quad
\Delta m_{31,32}^2/10^{-3}\mbox{eV}^2 =\left\{{{\begin{array}{*{20}c}
{\mbox{NH}:\quad2.517_{-0.028}^{+0.026}}\hfill \\
{\mbox{\,\,\,\!IH}:\!\!\!\!\!\quad-2.498_{-0.028}^{+0.028}}\hfill \\
\end{array}}}\right..
\end{align}
\label{eq2}
\end{subequations}

Since only the absolute values of $\Delta m_{31}^2$ and $\Delta m_{32}^2$ are known, it is possible to arrange the absolute values of the neutrino masses by two ways, namely, as a) $m_{1}<m_{2}<m_{3}$ and b) $m_{3}<m_{1}<m_{2}$. These two cases correspond to so called the normal ordering (NO) and the inverted ordering (IO) of the neutrino mass spectrum, respectively (in the last equation above the 31 subscript corresponds to the NO case and the 32 subscript corresponds to the IO case).

A few neutrino experiments at short distances (or, more exactly, at distances $L$, for which the numerical value of the $\Delta m_{ij}^2L/E$ parameter, with $E$ the neutrino energy, is of the order of unity) indicate the possible existence of sterile neutrinos with masses of the order of 1\,eV \cite{3,4}.
Such distances are considered as small ones and the experiments at such
distances are called as SBL (short-baseline) experiments. Light sterile
neutrinos are usually used to explain the SBL experimental data, which cannot be explained within the limited three-flavor mixing model with only massive active neutrinos. Firstly, it is the so called LSND/MiniBooNE or accelerator anomaly \cite{28,29}. Then refined calculations of the spectra of reactor antineutrinos were carried out \cite{30} that results in the higher calculated values of antineutrino fluxes, which do not confirm experimentally (see, however, the results presented in Ref.~\cite{31}). A possible deficit of reactor antineutrinos at short distances less than 100\,m is known now as the reactor (antineutrino) anomaly \cite{32}. A similar anomaly was observed also under calibration measurements for the experiments SAGE and GALLEX. This anomaly generally called as calibration or gallium one \cite{33,34,35}. The typical values of $\Delta m^{2}$ for the expected sterile neutrinos in all these cases are close to 1\,eV$^{2}$. Thus, these three types of observed anomalies for neutrinos fluxes (LSND/MiniBooNE or accelerator (AA),
gallium (GA), reactor (RA)) can be interpreted as evidences on the level of
about $3\sigma$ for the existence of light sterile neutrinos with masses of the order of 1\,eV. However, the additional experimental confirmation of these anomalies still requires. A list of such experiments can be found in
Ref.~\cite{7}. For example, in formation of flavor neutrino fluxes inside the Sun, the sterile neutrinos can be the reason for an exceed or a reduction of
an electron neutrinos yield in the transition region of neutrino energies about a few MeV. Results of relevant numerical calculations in the framework of the $(3+1+2)$-model are given below.

\section{The Main Statements of the Phenomenological (3+1+2)-Model of Neutrinos}

Let us consider the formalism of neutrino mixing in a generalized model of
neutrinos with three active neutrinos and three sterile neutrinos, that is
in the $(3+3)$-model \cite{14,14a,18,19,15}. More precisely, in the present paper, to take into account the effects of sterile neutrinos, the $(3+1+2)$-model is used \cite{14,14a,19}, which includes three known active neutrinos $\nu_a$ ($a=e,\mu,\tau$) and three new neutrinos (in this case light sterile ones): sterile neutrino $\nu_s$, hidden neutrino $\nu_h$ and dark neutrino $\nu_d$. In doing so it is suggested that two sterile neutrinos are degenerate with respect to their mass values as compared with a mass value of a third neutrino. Thus, the model contains six neutrino flavor states and six neutrino mass states, therefore a $6\!\times\!6$ mixing matrix should be used. This matrix is dubbed as the generalized mixing matrix or the generalized Pontecorvo--Maki--Nakagawa--Sakata matrix
$U_{\rm GPMNS}\equiv U_{\rm mix}$ \cite{19}. $U_{\rm mix}$ can be represented as the matrix product $V\!P$, where $P$ is a diagonal matrix containing the Majorana CP-phases $\phi_i$, $i=1,\dots,5$, that is
$P={\rm diag}\{e^{i\phi_1},\dots,e^{i\phi_5},1\}$. Below we will use only some particular forms of matrix $U_{\rm mix}$. In this case, we will denote the Dirac CP-phases as $\delta_i$ and $\kappa_j$, and the mixing angles as
$\theta_i$ and $\eta_j$. In doing so, $\delta_1\equiv\delta_{\rm CP}$,
$\theta_1\equiv\theta_{12}$, $\theta_2\equiv\theta_{23}$ and
$\theta_3\equiv\theta_{13}$. Only the normal ordering (NO) of the active
neutrino mass states and the value $\delta_{\rm CP}=1.1\pi$ will be considered. The results of the latest cosmological observations \cite{37a,37b,37c,37d, 37e,37f} justify the NO option fixed previously for mass states of active neutrinos with total sum of masses of active neutrinos about 0.06\,eV (see, for instance, \cite{14b}).

For compactness of formulas, we introduce symbols $\nu_b$ and $\nu_{i'}$ for
sterile left flavor fields and sterile left mass fields, respectively.
Fields $\nu_b$ with index $b$ contain fields $\nu_s$, $\nu_h$ and $\nu_d$,
while $i'$ denotes a set of indices $4$, $5$ and $6$. A total $6\!\times\!6$
mixing matrix $U_{\rm mix}$ can be represented in the form of $3\!\times\!3$
matrices $R$, $T$, $V$ and $W$:
\begin{equation}
\left(\begin{array}{c}\nu_a\\ \nu_b \end{array}\right)=
U_{\rm mix}\left(\begin{array}{c}\nu_i\\ \nu_{i'}\end{array}\right)\equiv
\left(\begin{array}{cc}R&T\\ V&W\end{array}\right)
\left(\begin{array}{c}\nu_i\\ \nu_{i'}\end{array}\right).
\label{eq_Umix}
\end{equation}
Let us represent the matrix $R$ in the form of $R=\varkappa U_{\rm PMNS}$,
where $\varkappa=1-\epsilon$, and $\epsilon$ is a small quantity. The matrix
$T$ in the equation~(\ref{eq_Umix}) must also be a matrix with small elements as compared with the matrix elements of the Pontecorvo--Maki--Nakagawa--Sakata $3\!\times\!3$ matrix for active neutrinos $U_{\rm PMNS}\equiv U$ ($UU^+=I$). So, active neutrinos mix by means of the matrix $U$, as it takes place in the $\nu$SM, with choosing the appropriate normalization. In the present state of the art, it is enough to restrict ourselves only to a few adjustable parameters of matrix $U_{\rm mix}$,
that allows one to interpret available (still rather heterogeneous) SBL
experimental data. The transition to the general matrix with full set of parameters should be done later on, if additional data related to the SBL anomalies will be obtained.

We choose $T$ in the form of $T=\sqrt{1-\varkappa^2}\,a$, where $a$ is an
arbitrary unitary $3\!\times\!3$ matrix ($aa^+=I$), then $U_{\rm mix}$ can be written in the following form:
\begin{equation}
U_{\rm mix}=\left(\begin{array}{cc}R&T\\ V&W\end{array}\right)\equiv
\left(\begin{array}{cc}\varkappa U&\sqrt{1-\varkappa^2}\,a\\
\sqrt{1-\varkappa^2}\,bU&\varkappa c \end{array}\right),
\label{eq_Utilde}
\end{equation}
where $b$ is also an arbitrary unitary $3\!\times\!3$ matrix ($bb^+=I$),
moreover $c=-ba$. Under these conditions the $U_{\rm mix}$ matrix will be
unitary, too ($U_{\rm mix}U_{\rm mix}^+=I$). In particular, we will use the
following $a$ and $b$ matrices:
\begin{subequations}
\begin{equation}
a=\left(\begin{array}{ccc}\cos\eta_1 \cos\eta_2 e^{-i\kappa_1} &
\cos\eta_1 \sin\eta_2 e^{-i\kappa_1} & \sin\eta_1 e^{-i\kappa_2}\\
-\sin\eta_2 & \cos\eta_2 & 0\\
-\sin\eta_1 \cos\eta_2 e^{-i\kappa_1}& -\sin\eta_1 \sin\eta_2 e^{-i\kappa_1} &
\cos\eta_1 e^{-i\kappa_2}\end{array}\right),
\label{eq_matricesa}
\end{equation}
\begin{equation}
b=-\left(\begin{array}{ccc}\cos\eta_1 \cos\eta_2 e^{i\kappa_1} & -\sin\eta_2
&-\sin\eta_1 \cos\eta_2 e^{i\kappa_1}\\
\cos\eta_1 \sin\eta_2 e^{i\kappa_1} & \cos\eta_2 &
-\sin\eta_1 \sin\eta_2 e^{i\kappa_1}\\
\sin\eta_1 e^{i\kappa_2} & 0 & \cos\eta_1 e^{i\kappa_2}\end{array}\right),
\label{eq_matricesb}
\end{equation}
\label{eq_matricesab}
\end{subequations}
where $\kappa_1$ and $\kappa_2$ are mixing phases between active and sterile
neutrinos, while $\eta_1$ and $\eta_2$ are mixing angles between them. The
remaining elements of the matrix $U_{\rm mix}$ are obtained in a standard way.

To make calculations more specific, we will use the following test values of
the new mixing parameters:
\begin{equation}
\kappa_1=\kappa_2=-\pi/2,\quad \eta_1=\pm 5^{\circ},\,\,\pm 15^{\circ},\,\,\pm 45^{\circ},
\quad \eta_2=\pm 30^{\circ},
\label{eq_etakappa}
\end{equation}
and restrict the values of the small parameter $\epsilon$ as
$\epsilon\lesssim 0.1$.

Let us specify neutrino masses by the set of values $\{m\}=\{m_i,m_{i'}\}$.
For active neutrino masses, we take the estimates presented in the
papers~\cite{14,14a,yud} for the NO case (in units of eV), which do not contradict recent experimental data: $m_1\approx 0.0016$,
$m_2\approx 0.0088$, $m_3\approx 0.0496$. The values of mixing angles $\theta_{ij}$ for three active neutrinos, which define Pontecorvo--Maki--Nakagawa--Sakata matrix, are calculated from the relations $\sin^2\theta_{12}\approx 0.304$, $\sin^2\theta_{23}\approx 0.573$ and $\sin^2\theta_{13}\approx 0.0222$. These relations are obtained on the basis of processing of experimental data for the NO-case and are given in Ref.~\cite{2}.

In order to choose the values of masses $m_4$ and $m_5$, we take into
account the results of the experiments BEST, DANSS, NEUTRINO-4, MiniBooNE and MacroBooNE \cite{35,ale,ser,agu22}. The BEST, DANSS and NEUTRINO-4 experiments are devoted to testing the existence of GA and RA associated with a deficit of electron neutrinos and antineutrinos, respectively. It would be expected that the values of the sterile neutrino mass measured in these experiments are practically the same. The value of $m_4=1.1$ in eV agrees with the data of the BEST experiment and coincides with the value of this parameter previously used in our $(3+3)$ model \cite{19}. For the value of $m_5$, we take either the value $0.6$~eV close to the results of the MiniBooNE and MacroBooNE experiments \cite{agu22}, or $0.003$~eV \cite{14a}.  As for the value of $m_6$, its justified choice can be made according to the results of future special experiments. Nonetheless we choose $m_6$ equal to $0.001$~eV following Ref.~\cite{14a}. Thus, in this paper we consider two cases for mass values $m_4$, $m_5$ and $m_6$ in eV: the A case (two comparatively heavy states and one light state with $m_4=1.1$, $m_5=0.6$, $m_6=0.001$ and $\eta_2=\pi/6$) and the B case (one comparatively heavy state and two light  states with $m_4=0.45$, $m_5=0.003$, $m_6=0.001$ and $\eta_2=\pi/12$). The A case uses the $m_4$ and $m_5$ mass values, which are close to results based on experimental data from Refs.~\cite{sinev,aleks}. The B case uses the $m_4$, $m_5$ and $m_6$ mass values, which are close to ones from the Refs.~\cite{agu22,14a}.

As is known, the problem of the origin of the masses of fundamental fermions, which are strongly various with their values, is still not completely understood. In the framework of the $\nu$SM, these values arise due to Yukawa couplings between the fields of the neutrinos and the Higgs field. However, the values of the neutrino masses are very small, so probably a mechanism of their appearance is basically related to the Majorana nature of neutrinos. The problem of neutrino masses was considered at a phenomenological level by many authors \cite{14,14a,15,40,41,42,43}. For example, in Ref.~\cite{14a} the possible correlation between the scale of the minimal neutrino mass and the linear scale of the Universe dark energy density \cite{21,22} had been supposed.

Using known equations for propagation of various neutrino flavors (see, for
example,~\cite{44}), it is possible to obtain analytical expressions for the
transition probabilities of various flavors of stable neutrinos/antineutrinos in a vacuum as a function of distance $L$ from the source. Besides it is
possible to generalize analytical expressions for the probabilities of
transitions and conservation of various neutrino flavors \cite{Bilenky} to the case of decaying neutrinos with including of their decay widths $\Gamma_k$ \cite{19}. To do this, it is necessary to substitute $E_k-i\Gamma_k/2$ instead of the neutrino energy $E_k$ into the original
equations for the propagation of neutrino flavors, where $\Gamma_k\approx m_k\Gamma_k^{(0)}/E_k$ is the decay width in the laboratory frame, and $\Gamma_k^{(0)}$ is the same in the rest frame. It seems plausible that results of both carried out and planned experiments will lead to necessity of taking into account of sterile neutrinos decay widths \cite{19}. In this instance $\nu_4$ and $\nu_5$ may have nonzero $\Gamma_4$ and $\Gamma_5$ values and corresponding decay channels, which produce $\nu_6$ in a final state, while obviously $\nu_6$ has not decay channels. So $\nu_6$ and practically $\nu_d$ could be stable components of the fermionic dark matter.

\section{Oscillations of Active and Sterile Neutrinos Inside the Sun}

The probability amplitudes for propagation of neutrino flavors can be found by solution of well-known equations (see, for example, \cite{44,14a,yud}).
For three active neutrinos as ultrarelativistic particles these equations have the form
\begin{equation}
i\partial_{r}\left(\begin{array}{c} a_{e}\\ a_{\mu}\\ a_{\tau}\end{array}
\right)=H\left(\begin{array}{c} a_{e}\\ a_{\mu}\\
a_{\tau}\end{array}\right),
\label{eq_active_n_amplitudes}
\end{equation}
where the matrix $H$ is expressed with using the matrix $U_{\rm PMNS}\equiv U$ in the form of
\begin{equation}
H=\frac{U}{2E}\!\left(\begin{array}{ccc}m_{1}^{2}-m_{0}^{2} & 0 & 0 \\
0 & m_{2}^{2}-m_{0}^{2} & 0 \\ 0 & 0 & m_{3}^{2}-m_{0}^{2}
\end{array}\right)\!U^{+}.
\label{eq_activeH}
\end{equation}
Here $m_0$ is the smallest value among three neutrino masses $m_1$, $m_2$ and $m_3$, and $E$ is the neutrino energy. In what follows, as a basic case it
will be used here the simplest conventional approach for neutrino oscillations that is based on the plane-wave neutrino states.
In the plane-wave approximation the neutrinos possess equal momentums that
leads to the diagonal energy matrix $\Delta_{E}$ in the following form:
\begin{equation}
\Delta_{E}={\rm diag}\{E_1-E_0,\,E_2-E_0,\,\ldots,\,E_{6}-E_0\}\,,
\label{eq_DeltaE}
\end{equation}
where $E_i=\sqrt{p^2+m_i^2}$, $m_i$ ($i=1,2,\ldots 6$) are the neutrino masses and $m_0$ is the smallest mass among $m_i$. The momentum $p$ can be
related to the energy $E\approx p$ of ultrarelativistic active neutrinos. In
the ultrarelativistic limit for all neutrinos, instead of the matrix
$\Delta_{E}$ it is possible to use the matrix $\Delta_{m^2}$ of the differences of the squares of masses, which in the general case for 3+$N$ flavors is defined as
\begin{equation}
\Delta_{m^2}={\rm diag}\{m_1^2-m_0^2,\,m_2^2-m_0^2,\,\ldots,
\,m_{3+N}^2-m_0^2\}\,.
\label{eq_minsquarematrix}
\end{equation}
Then it is necessary to solve the following equations for neutrino
propagation, similar to the equations~(\ref{eq_active_n_amplitudes}) and
(\ref{eq_activeH}) for active neutrinos:
\begin{equation}
i\partial_{r}\left(\begin{array}{c} a_{a}\\ a_{b} \end{array}\right)=
\frac{U_{\rm mix}}{2E}\Delta_{m^2}U^{+}_{\rm mix}
\left(\begin{array}{c} a_{a}\\ a_{b} \end{array}\right),
\label{eq_neutrino_m2}
\end{equation}
where $U_{\rm mix}$ is the unitary $6\!\times\!6$ neutrino mixing matrix given by equations (\ref{eq_Utilde})--(\ref{eq_matricesab}) and $r=r_a\approx ct$ is the distance travelled by active neutrinos (it was assumed above that $c=1$). For antineutrinos, the equations have the form
\begin{equation}
i\partial_{r}\left(\begin{array}{c} {a}_{\overline{a}}\\ {a}_{\overline{b}}
\end{array}\right)=
\frac{U_{\rm mix}^*}{2E}\Delta_{m^2}U^T_{\rm mix}\left(\begin{array}{c}
{a}_{\overline{a}}\\ {a}_{\overline{b}} \end{array}
\right),
\label{eq_antineutrino_m2}
\end{equation}
where $\ast$ means complex conjugation. Solving these equations for given
values of the parameters, one can find the survival probabilities and also
probabilities of appearance and disappearance of neutrinos or antineutrinos
of any flavor as functions of the neutrino (or antineutrino) energy and the
distance from the source.

The well-known equations for the amplitudes of the neutrino propagating in the medium in the case of three active neutrinos are as follows (see, for example, \cite{yud}):
\begin{equation}
i\partial_{r}\!\left(\!\begin{array}{c}
a_{e}\\ a_{\mu}\\ a_{\tau}
\end{array}\!\right)=\left\{\!\frac{U}{2E}\!
\left(\!\begin{array}{ccc}
m_{1}^{2}-m_{0}^{2} & 0 & 0 \\
0 & m_{2}^{2}-m_{0}^{2} & 0 \\
0 & 0 & m_{3}^{2}-m_{0}^{2}
\end{array}
\!\right)\!U^{+}\right.+\left.\left(\begin{array}{ccc}
\sqrt{2}G_{F}N_{e}(r) & 0 & 0 \\
0 & 0 & 0 \\
0 & 0 & 0
\end{array}
\right)\right\}\left(\!\begin{array}{c}
a_{e}\\ a_{\mu}\\ a_{\tau}
\end{array}\!\right),
\label{eq11}
\end{equation}
where  $N_{e}(r)$ is the electron density in the medium, where neutrinos
propagate, and $G_F$ is a Fermi constant.

Generally, when the number of different types of neutrinos is $3+N$, via the
mixing matrix $U$ we can define a matrix $\tilde{\Delta}_{m^2}=
\tilde{U}\Delta_{m^2}\tilde{U}^{+}$, with $\Delta_{m^2}
={\rm diag}\{m_1^2-m_0^2,m_2^2-m_0^2,\ldots,m_{3+N}^2-m_0^2\}$. In the case of the $(3+3)$-model, we obtain the following equation replacing Eq.~(\ref{eq11}):
\begin{equation}
i\partial_{r}\!\left(\!\begin{array}{c}
a_{\alpha}\\ a_{b} \end{array}\!\right)
=\left[\frac{\tilde{\Delta}_{m^2}}{2E}
+\sqrt{2}G_{F}\!\left(\!\begin{array}{cc}
\tilde{N}_{e}(r) & 0 \\
0  & \tilde{N}_{n}(r)/2\end{array}\!\right)\right]
\!\!\left(\!\begin{array}{c}
a_{\alpha}\\ a_{b}\end{array}\!\right)\!,
\label{eq12}
\end{equation}
where $\tilde{N}_e(r)$ and $\tilde{N}_n(r)$ are the $3\times3$-matrices
presented below:
\begin{equation}
\tilde{N}_e(r)={N_e(r)}\left({{\begin{array}{*{20}c}
1\hfill & 0\hfill & 0\hfill \\
0\hfill & 0\hfill & 0\hfill \\
0\hfill & 0\hfill & 0\hfill \\
\end{array}}}\right), \quad
\tilde{N}_n(r)=N_n(r)\left({{\begin{array}{*{20}c}
1\hfill & 0\hfill & 0\hfill \\
0\hfill & 1\hfill & 0\hfill \\
0\hfill & 0\hfill & 1\hfill \\
\end{array}}}\right).
\label{eq13}
\end{equation}

\begin{figure}[htbp]
\centering
\includegraphics[width=0.99\textwidth]{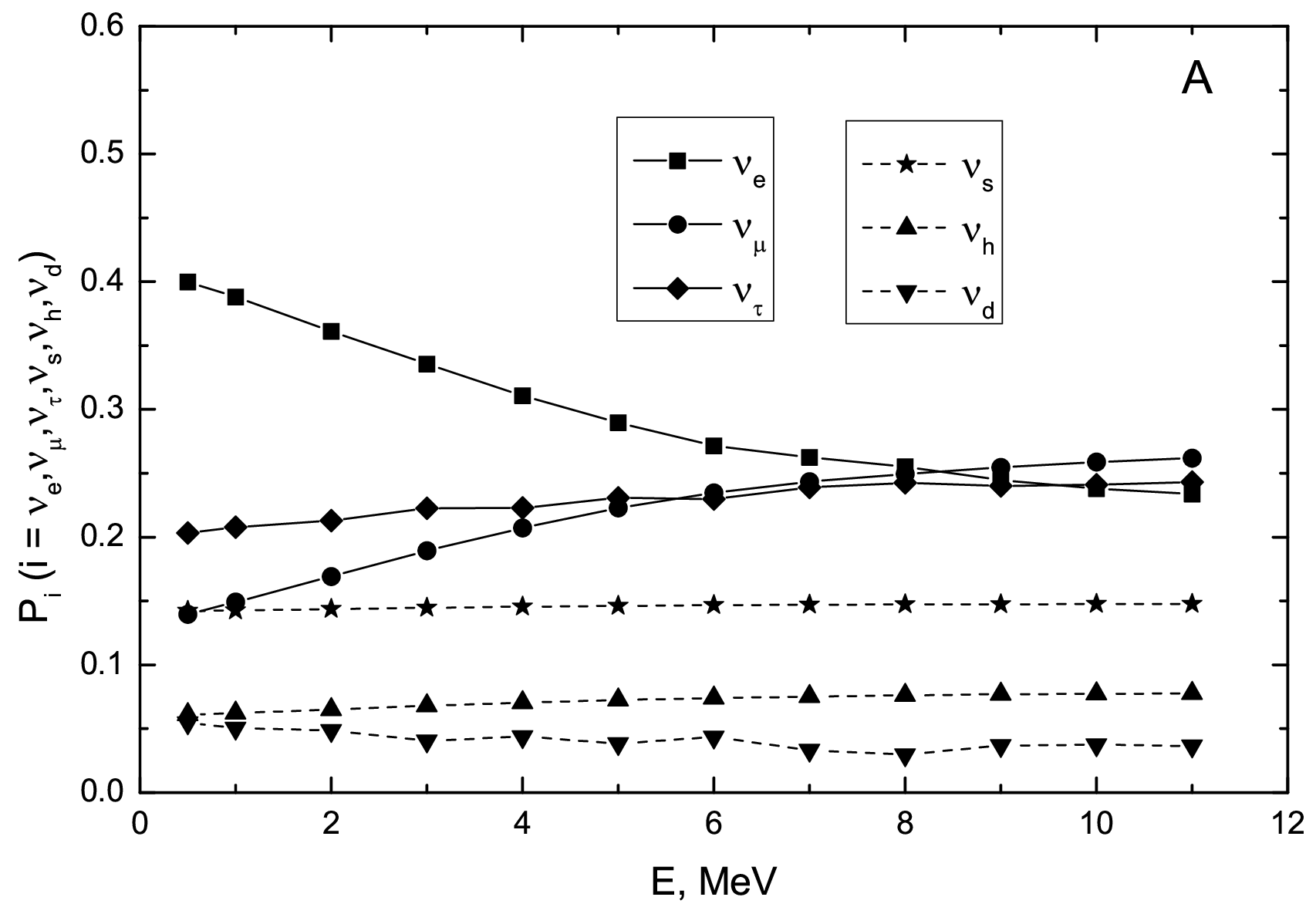}
\caption{The mean probabilities $P_i$ for the A case of the yield of three
different types of active neutrinos (solid curves) and of three different
types of sterile neutrinos (dashed curves) on the Sun surface versus the
neutrino energy in the $(3+1+2)$-model of both active and sterile neutrinos
with the parameters $m_4=1.1$~eV, $m_5=0.6$~eV,
$m_6=0.001$~eV, $\eta_1=\pi/12$, $\eta_2=\pi/6$, $\epsilon=0.08$.}
\label{fig1}
\end{figure}
\begin{figure}[htbp]
\centering
\includegraphics[width=0.99\textwidth]{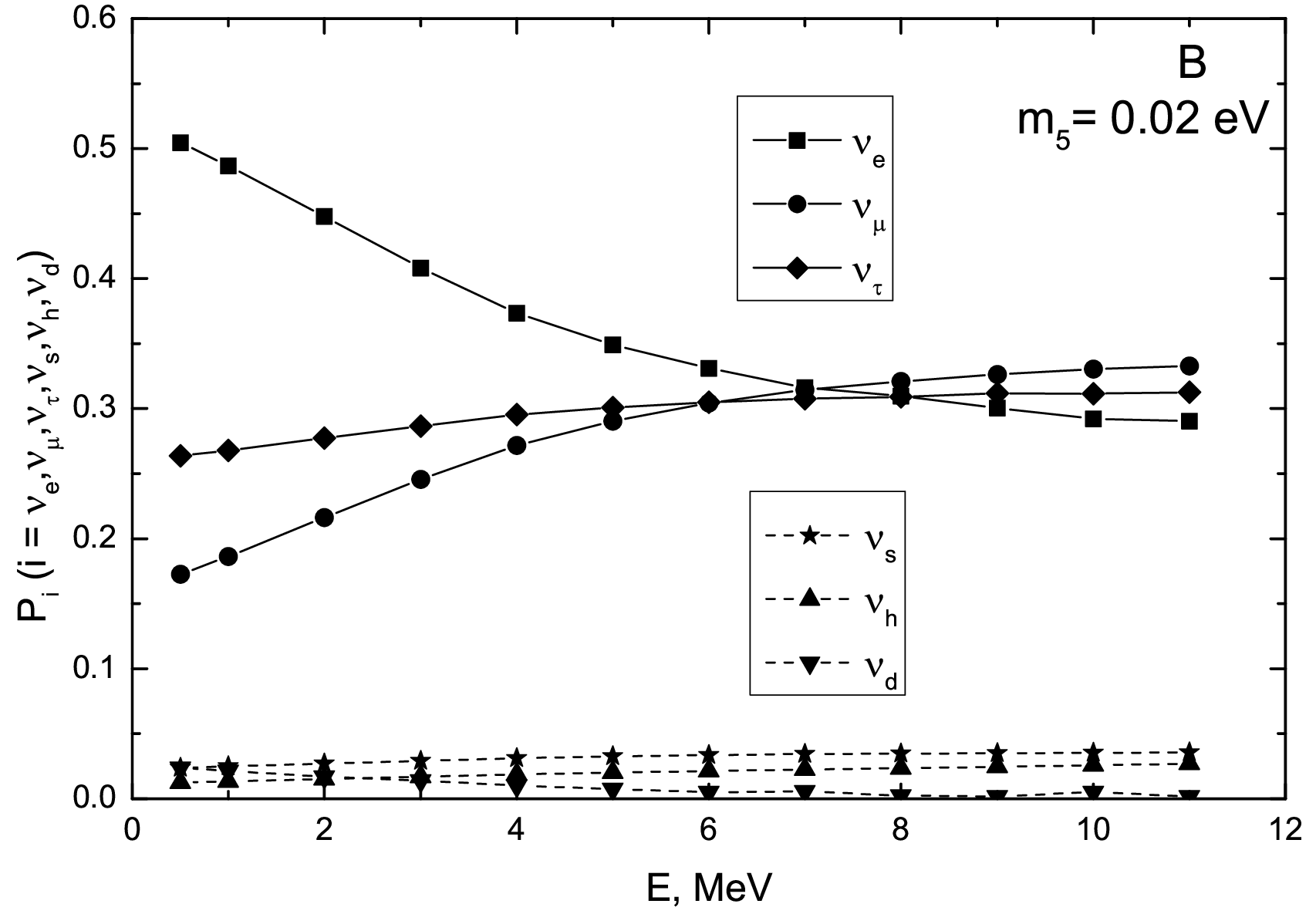}
\caption{The mean probabilities $P_i$ for the B case of the yield of three
different types of active neutrinos (solid curves) and of three different
types of sterile neutrinos (dashed curves) on the Sun surface versus the
neutrino energy in the $(3+1+2)$-model of both active and sterile neutrinos
with the parameters $m_4=0.45$~eV, $m_5=0.02$~eV,
$m_6=0.001$~eV, $\eta_1=\pi/4$, $\eta_2=\pi/6$, $\epsilon=0.02$.}
\label{fig2}
\end{figure}

\begin{figure}[htbp]
\centering
\includegraphics[width=0.99\textwidth]{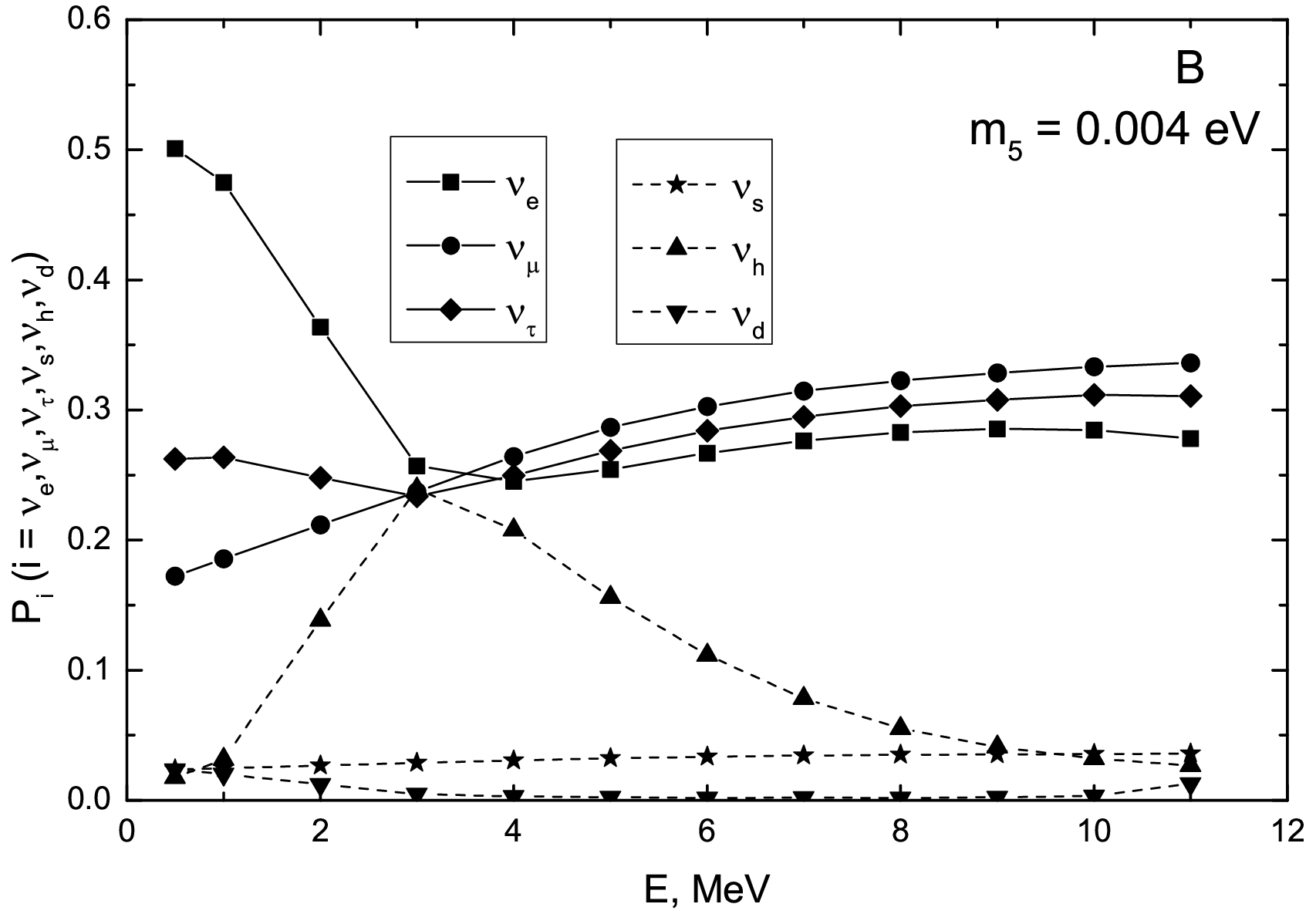}
\caption{The mean probabilities $P_i$ for the B case of the yield of three
different types of active neutrinos (solid curves) and of three different
types of sterile neutrinos (dashed curves) on the Sun surface versus the
neutrino energy in the $(3+1+2)$-model of both active and sterile neutrinos
with the parameters $m_4=0.45$~eV, $m_5=0.004$~eV,
$m_6=0.001$~eV, $\eta_1=\pi/4$, $\eta_2=\pi/6$, $\epsilon=0.02$.}
\label{fig3}
\end{figure}

\begin{figure}[htbp]
\centering
\includegraphics[width=0.99\textwidth]{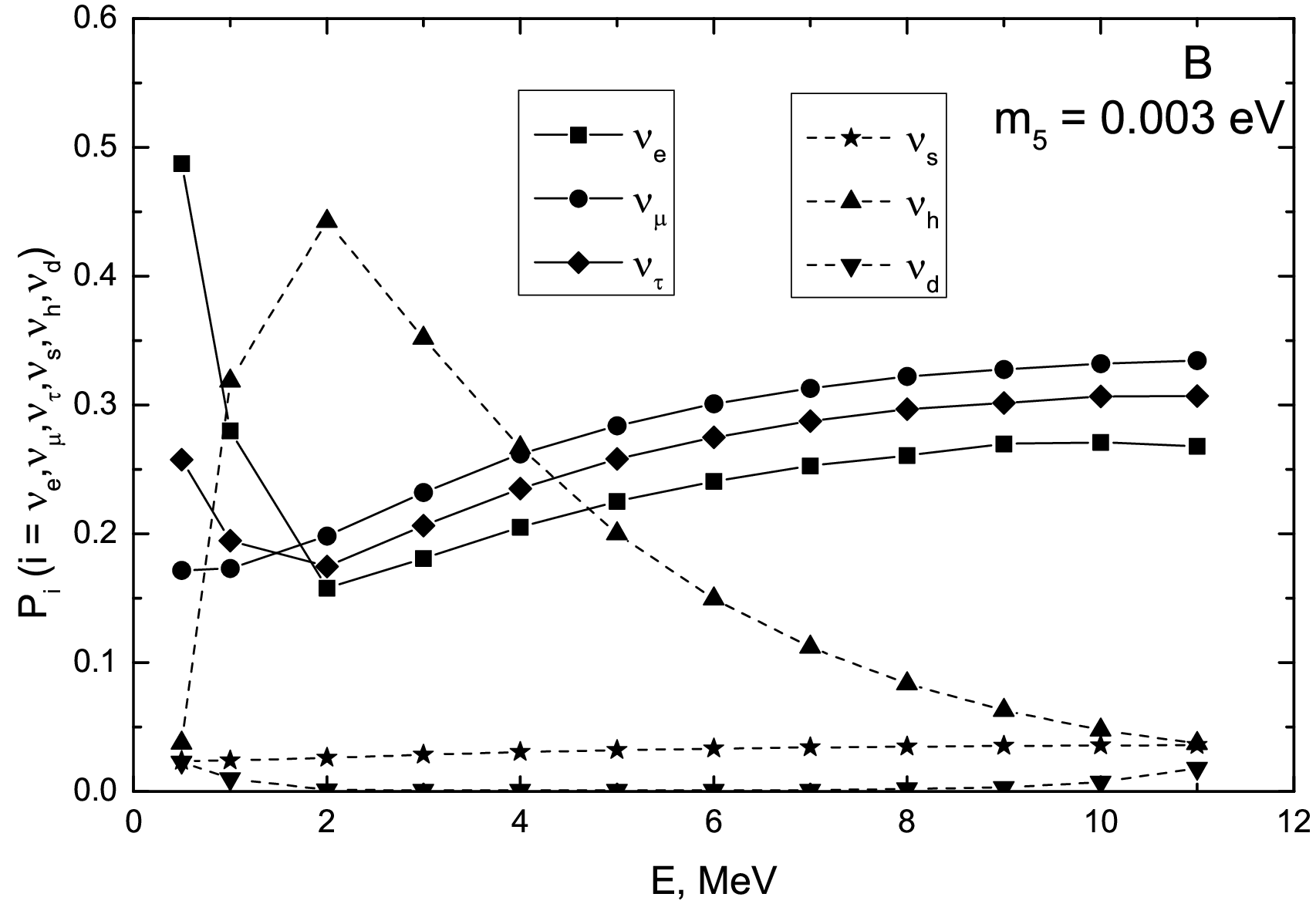}
\caption{The mean probabilities $P_i$ for the B case of the yield of three
different types of active neutrinos (solid curves) and of three different
types of sterile neutrinos (dashed curves) on the Sun surface versus the
neutrino energy in the $(3+1+2)$-model of both active and sterile neutrinos
with the parameters $m_4=0.45$~eV, $m_5=0.003$~eV,
$m_6=0.001$~eV, $\eta_1=\pi/4$, $\eta_2=\pi/6$, $\epsilon=0.02$.}
\label{fig4}
\end{figure}

Let us solve Eq.~(\ref{eq12}) inside the Sun with the use of the electron
density $N_e(r)$ and the neutron density $N_n(r)$ obtained in the standard
solar model (SSM) \cite{45,46,47}. Then it is performed the local averaging in each point of the solutions obtained, which oscillate very rapidly as a function of the radial variable $r$, and we can find the smooth averaged survival probability of the electron neutrino and the appearance probability of another neutrino flavors against the distance $r$ from the center of the Sun for different neutrino energies. Of main interest here are the mean  probabilities $P_i$ ($i=e,\mu,\tau,s,h,d$) of the neutrino yields for different flavors, both active and sterile, just on the surface of the Sun, which are shown in Figs.~\ref{fig1}--\ref{fig4}. For example, Figs.~\ref{fig1} and \ref{fig2} show the neutrino energy dependence of the electron neutrino survival probability and the appearance probabilities of muon, tau and sterile neutrinos in the energy range from 0.5 to 11\,MeV for the A-case and the B-case at $m_{5}=0.02$~eV, respectively. In this energy range one can see a monotonous behaviour for all probabilities. On Fig.~\ref{fig3} the energy dependences for these probabilities are shown for the B-case at $m_{5}=0.004$~eV. The behaviour for the appearance probability of muon, sterile and dark neutrinos does not change practically, while the energy dependence of the electron neutrino survival probability and the appearance probability of tau and hidden neutrinos essentially change at energy about 3\,MeV. The dips for electron and tau neutrinos are seen, while the growth appears for hidden neutrinos in this range. On Fig.~\ref{fig4}, the dependences for these probabilities are shown for the B-case at $m_{5}=0.003$~eV. It is seen that the strong resonance for hidden neutrinos is formed at neutrino energy about 2\,MeV with the corresponding
responses for electron and tau neutrinos. So Figs.~\ref{fig3} and \ref{fig4}
show the resonance behaviour of hidden neutrinos with mass $\approx 0.003$~eV when electron neutrinos oscillate and interact with the Sun matter at neutrino energy about 2\,MeV. This characteristic property of the considered model can be used at its investigation and possible further applications.


\section{Conclusion}

Properties of neutrinos are rather mysterious and further intensive theoretical and experimental studies are required to determine the nature and characteristics of these unusual elementary particles. Construction and
development of adequate phenomenological models of neutrinos, which generalize the SM in the neutrino sector, is one of the ways for the interpretation of new experimental results and also searching the definite GUT. It is therefore of great interest to verify existence and properties of sterile neutrinos and to determine their number and absolute mass scales for both active and sterile neutrinos, also including for this goal  investigations of the values of solar neutrinos fluxes with different energies \cite{gold,gon,bor}.

In this paper, the phenomenological $(3+1+2)$-model was used to demonstrate
the properties of the neutrino oscillations inside the Sun among the three
active and three sterile neutrinos. While considering the oscillations of
active and sterile neutrinos in the solar medium, the density profiles of the electrons and neutrons obtained in the SSM \cite{45,46,47} were used. The
calculations at chosen parameters values show the resonance effect in the
solar medium for the hidden (fifth )neutrino with its mass value $m_5$ about $0.003$~eV at neutrino energies near 2\,MeV. This effect can be a characteristic feature of the considered model at its verification and possible application. Note that chosen values of the parameters of the $(3+1+2)$-model in sterile neutrino sector are very conditional and were used mainly for illustration of manifesting effects. Detail investigations
with more precise parameters values will be done elsewhere within the
considered model on the basis of new data from the sterile neutrinos search
experiments and solar neutrino observations.

\section{Acknowledgements}

The authors are grateful to M.~D. Skorokhvatov, Yu.~S. Lyutostansky,
A.~G. Doroshkevich and V.~I. Lyashuk for useful discussions.

\end{document}